\documentclass[conference, letterpaper]{IEEEtran}
\usepackage{cite}
\usepackage{amsmath,amssymb,amsfonts}
\usepackage{algorithmic}
\usepackage{graphicx}
\usepackage{textcomp}
\usepackage{xcolor}
\def\BibTeX{{\rm B\kern-.05em{\sc i\kern-.025em b}\kern-.08em
    T\kern-.1667em\lower.7ex\hbox{E}\kern-.125emX}}

\usepackage{tabularx} % for uniform column width

\usepackage{multirow}

\usepackage{geometry}
\usepackage[numbers]{natbib} %included for arXiv %%%%%

\usepackage[hidelinks]{hyperref} %for arXiv for breaking long urls
\usepackage{xurl} %used xurl instead of url to line break long urls in references

\begin{document}

\title {Standardised schema and taxonomy for AI incident databases in critical digital infrastructure\\
}

\author{\IEEEauthorblockN{Avinash Agarwal}
\IEEEauthorblockA{\textit{Telecommunication Engineering Centre} \\
\textit{Ministry of Communications}\\
New Delhi, India \\
avinash.70@gov.in \\
https://orcid.org/0000-0003-4553-5861}
\and
\IEEEauthorblockN{Manisha J. Nene}
\IEEEauthorblockA{\textit{Defence Institute of Advanced Technology} \\
\textit{Ministry of Defence}\\
Pune, India \\
mjnene@diat.ac.in \\
https://orcid.org/0000-0003-0669-4464}
}

\maketitle

\begin{abstract}
The rapid deployment of Artificial Intelligence (AI) in critical digital infrastructure introduces significant risks, necessitating a robust framework for systematically collecting AI incident data to prevent future incidents. Existing databases lack the granularity as well as the standardized structure required for consistent data collection and analysis, impeding effective incident management. This work proposes a standardized schema and taxonomy for AI incident databases, addressing these challenges by enabling detailed and structured documentation of AI incidents across sectors. Key contributions include developing a unified schema, introducing new fields such as incident severity, causes, and harms caused, and proposing a taxonomy for classifying AI incidents in critical digital infrastructure. The proposed solution facilitates more effective incident data collection and analysis, thus supporting evidence-based policymaking, enhancing industry safety measures, and promoting transparency. This work lays the foundation for a coordinated global response to AI incidents, ensuring trust, safety, and accountability in using AI across regions.
\end{abstract}


\section{Introduction}
\label{introduction}
Increasing integration of AI systems into critical digital infrastructure offers both transformative benefits and significant risks. The potential for AI-driven failures or unforeseen consequences necessitates the establishment of robust incident reporting mechanisms. A standardized AI incident reporting mechanism would allow for the systematic collection and detailed analysis of incident data, helping to uncover trends, identify vulnerabilities, and prevent future incidents by learning from past incidents \citep{mcgregor2021preventing}. Such knowledge is essential for devising effective mitigation strategies, developing safety protocols, evolving considered best practices, and informing regulatory frameworks that govern the use of AI in these sensitive domains \citep{turri2023we}. Recent regulatory developments further underscore the need for such a framework. These include the EU AI Act, which mandates reporting of serious incidents involving high-risk AI systems, including those in critical infrastructure \citep{EU2024AIact}. To quote the EU AI Act, \emph{serious incident means an incident or malfunctioning of an AI system that directly or indirectly leads to any of the following: (a) the death of a person, or serious harm to a person’s health; (b) a serious and irreversible disruption of the management or operation of critical infrastructure. (c) the infringement of obligations under Union law intended to protect fundamental rights; (d) serious harm to property or the environment.}

Despite the pressing need, a standardized approach for documenting and analyzing AI incidents in these sectors is currently lacking \citep{lupo2023risky}. The effectiveness of current AI incident databases, such as the Artificial Intelligence Incident Database (AIID) \citep{mcgregor2021ai} and the AI, Algorithmic, and Autonomous Incident Classification (AIAAIC) Repository \citep{AIAAICRepository}, is hampered by several key issues. These databases vary significantly in their structure and the granularity of the data they capture \citep{turri2023we}, making it complicated to aggregate and analyze information across different sectors and regions. For instance, while some databases focus on high-level incident descriptions, others lack detailed fields necessary for thorough analysis. Additionally, the absence of a unified taxonomy for categorizing incidents further complicates efforts to draw actionable insights from the data \citep{10677312}. These inconsistencies highlight the need for a more standardized approach to AI incident reporting that can ensure comprehensive, comparable, and actionable data \citep{agarwal2024advancing}.

This study addresses these challenges by developing a standardized schema and taxonomy for AI incident databases. The objective is to create a unified framework that can be adopted across various sectors, enabling consistent and detailed documentation of AI incidents. By standardizing the AI incident reporting and categorization methodology, this research aims to enhance the quality and utility of AI incident data, providing a stronger foundation for analyzing risks, developing mitigation strategies, and informing policy decisions.

This paper makes two significant contributions. First, it establishes a standardized schema for AI incident reporting, which enhances the granularity and consistency of data collection across various databases. Second, it introduces a taxonomy for classifying AI incidents in critical digital infrastructure, improving the comprehensiveness and clarity of incident data.

The structure of this paper is as follows: Section \ref{analysis} analyses existing AI incident databases and highlights their structural gaps. Section \ref{results} presents the results of this work, including the proposed standardized schema and taxonomy. Section \ref{discussion} discusses the implications of the schema and dataset, while Section \ref{conclusion} concludes with a summary of findings and recommendations for future research.

\section{Analysis of existing AI incident databases}
\label{analysis}
This section reviews existing AI incident databases worldwide, highlighting their limitations in systematically collecting and categorizing incident data.

\subsection{AIAAIC Repository}
The AIAAIC Repository \citep{AIAAICRepository}, which stands for 'AI, Algorithmic, and Automation Incidents and Controversies', is an evolving open resource that documents incidents and controversies related to artificial intelligence, algorithms, and automation. It collects and classifies these occurrences, shedding light on the ethical, social, and technical dimensions that drive them. AIAAIC emphasizes broader impacts such as job displacements, environmental damage, and misleading marketing. The repository covers a wide array of issues, with 1009 incidents and 411 issues reported as of September 2024. This resource is publicly accessible, aiming to serve as a tool for researchers, policymakers, educators, and the general public to understand and navigate the complexities of AI and automation.

\subsection{AI Incident Database (AIID)}
The AIID \citep{AIIncidentDatabase} is another resource dedicated to documenting AI-related incidents. Inspired by similar incident databases in aviation and computer security, AIID seeks to record past AI incidents. It compiles real-world harms or near harms caused by AI systems across various domains, making it a non-domain-specific resource. As of September 2024, AIID has recorded 759 incidents and over 3,500 reports. Managed by an industrial/non-profit cooperative, the database is an open resource accessible to the public, with incident submissions reviewed before being added. It also encourages community participation, recognizing that collective input is crucial for improving AI safety and accountability \citep{mcgregor2021preventing}. 

\subsection{Other AI incident repositories}
In addition to these two AI incident databases, there are a few other repositories for documenting AI-related incidents. The AI Vulnerability Database (AVID) \citep{AVIdatabase} catalogs about 40 AI system vulnerabilities, including general failure modes and specific reports of these failures. AVID uses a detailed taxonomy to categorize issues, such as security, ethics, and performance, providing AI developers and auditors with valuable evaluation methods.

The AI Incidents Monitor (AIM) is a developing initiative by the OECD.AI expert group designed to track real-time AI incidents \citep{OECDAIM}. AIM aims to provide an evidence base to inform AI incident reporting frameworks and related policy discussions. Unlike AIID and AIAAIC, AIM does not currently allow open submissions.

Another repository, 'Where in the World is AI?' presents AI incidents on an interactive global map, highlighting responsible or unethical uses of AI globally. It emphasizes the geographical context, labeling incidents as either harmful or helpful. However, the map has not been updated since 2021, and its database is currently inaccessible.

In addition to these, there are informal lists on platforms like Twitter and GitHub that chronicle problematic AI systems. These lists lack formal taxonomies and reporting structures but still represent early efforts to document irresponsible AI use. 

Lastly, various government databases, such as those in the European Union, Amsterdam, Helsinki, and Chile, provide transparency by cataloging AI systems used in the public sector. These registers offer detailed information about the designs, contexts, and impacts of various AI systems, serving as useful tools for public awareness and governance.

\subsection{Database Schema Discrepancies}
Table \ref{table1} presents the data fields of the two databases, highlighting their vastly different and incompatible structures. Additionally, both databases lack fields required for capturing detailed structured information necessary for a thorough analysis of incidents, such as the causes, context, and impacts. For illustration, AIAAIC does not have fields for the affected parties and incident summaries. On the other hand, AIID does not have fields for the concerned application name, its technology and purpose, impacted sectors, and so on.

%%%%%%%%%%
\begin{table}[h]
\caption{Data fields available in AIAAIC and AIID}
\label{table1}
\setlength{\tabcolsep}{2pt}  % reduces spaces in cell padding?
\begin{center}
\begin{tabularx} {\columnwidth} {|
>{\raggedright\arraybackslash}>{\hsize=1\hsize} X |
>{\raggedright\arraybackslash}>{\hsize=1\hsize} X |
}
\hline
  \textbf{Fields available in AIAAIC} &
  \textbf{Fields available in AIID} \\ \hline
\begin{tabular}[X]{@{}X@{}}AIAAIC ID;\\ Headline/title;\\ Type;\\ Released;\\ Occurred;\\ Country(ies);\\ Sector(s);\\ Deployer(s);\\ Developer(s);\\ System name(s);\\ Technology(ies);\\ Purpose(s);\\ Media trigger(s);\\ Issue(s);\\ Transparency;\\ Harm(s)*\\ -External harms (Individual, Societal, Environmental);\\ - Internal harms (Strategic/ reputational, Operational, Financial, Legal/regulatory);\\ Description/links\\ \emph{*Harm data is only accessible to Premium Members} \end{tabular} &  \begin{tabular}[X]{@{}X@{}}Incident-id;\\ Title;\\ Description;\\ Date;\\ Alleged deployer of AI system;\\ Alleged developer of AI system;\\ Alleged harmed or nearly harmed parties\\ \end{tabular} \\ \hline
\end{tabularx}
\end{center}
\end{table}
%%%%%%%%%%

 \subsection{Inconsistent definitions and taxonomies}
Another challenge is the lack of standardized definitions and taxonomies for classifying AI incidents. Each database uses its own criteria, which may not align with regulatory definitions. For example, AIAAIC incident with id AIAAIC1724 \citep{AIAAIC1724} pertains to smog caused by gas turbines of a data center of xAI company. This incident is not an outcome of any AI application, and recording it as an AI incident is not accurate. The same is the case with incident id AIAAIC1695 \citep{AIAAIC1695}, related to industrial waste dumped by Microsoft's Mekaguda data center, which also should not be listed as an AI incident. Similarly, incident id AIAAIC1561 \citep{AIAAIC1561} relates to a lawsuit filed against OpenAI and Microsoft for alleged copyright infringements while training their AI models, which may not fit into the OECD's definition of an AI incident \citep{OECDAIIncident}.

The absence of a consistent framework further complicates the evaluation of incident severity. Both databases do not provide detailed classifications for types of harm, their severity, and their root causes. They also lack specific information on AI applications, such as version numbers. Without standardized definitions and taxonomies, it is challenging to categorize incidents effectively, identify commonalities, and analyze patterns across different databases.

In conclusion, there are only a few AI incident databases available globally. Further, they suffer from several significant shortcomings. These include insufficient data fields, incompatible schemas, and a lack of standardized taxonomies. Also, it is challenging to classify incidents for systematic analysis due to varying definitions of an AI incident and a lack of detailed data. Additionally, these databases are often general-purpose, with limited attention to specific sectors, such as critical digital infrastructure, where more detailed and sector-specific information is crucial. These limitations hinder the effectiveness of these databases in providing comprehensive insights into AI-related incidents. This work addresses these gaps by proposing a standardized schema and taxonomy.

\section{Results}
\label{results}
The results section presents two key outcomes: the developed standardized schema and the proposed taxonomy for AI incident databases for critical digital infrastructure.

\subsection{Proposed standardized schema}
In developing a standardized schema for AI incident reporting, existing fields from the AIID, AIAAIC, and other repositories were carefully reviewed and incorporated. Further, additional fields were introduced to enhance the comprehensiveness and utility of the database. These additions, such as incident causes, incident severity, and distinct types of harm, enable a deeper understanding of the root causes and diverse impacts of AI incidents. Table \ref{table2} presents the proposed standardized fields of an AI incident database.
%%%%%%%%
\begin{table}[h]
\caption{Proposed standardized schema of an AI incident database}
\label{table2}
\begin{center}
\begin{tabular}{|lcc|}
\hline
\multicolumn{1}{|l|}{\textbf{Standardised fields}}               & \multicolumn{1}{c|}{\textbf{AIID}} & \textbf{AIAAIC} \\ \hline
\multicolumn{1}{|l|}{Incident ID}              & \multicolumn{1}{c|}{Yes} & Yes \\ \hline
\multicolumn{1}{|l|}{Incident title}           & \multicolumn{1}{c|}{Yes} & Yes \\ \hline
\multicolumn{1}{|l|}{Incident summary}         & \multicolumn{1}{c|}{Yes} & No  \\ \hline
\multicolumn{1}{|l|}{Incident date}            & \multicolumn{1}{c|}{Yes} & Yes \\ \hline
\multicolumn{1}{|l|}{Incident location(s)}     & \multicolumn{1}{c|}{No}  & Yes \\ \hline
\multicolumn{1}{|l|}{Affected party(ies)}      & \multicolumn{1}{c|}{Yes} & No  \\ \hline
\multicolumn{1}{|l|}{Sector(s) impacted}       & \multicolumn{1}{c|}{No}  & Yes \\ \hline
\multicolumn{1}{|l|}{Incident issue(s)}        & \multicolumn{1}{c|}{No}  & Yes \\ \hline
\multicolumn{1}{|l|}{AI application name(s)}   & \multicolumn{1}{c|}{No}  & Yes \\ \hline
\multicolumn{1}{|l|}{Application version}      & \multicolumn{1}{c|}{No}  & No  \\ \hline
\multicolumn{1}{|l|}{Application technology(ies)}                & \multicolumn{1}{c|}{No}            & Yes             \\ \hline
\multicolumn{1}{|l|}{Application purpose(s)}   & \multicolumn{1}{c|}{No}  & Yes \\ \hline
\multicolumn{1}{|l|}{Application deployer}     & \multicolumn{1}{c|}{Yes} & Yes \\ \hline
\multicolumn{1}{|l|}{Application developer}    & \multicolumn{1}{c|}{Yes} & Yes \\ \hline
\multicolumn{1}{|l|}{Application transparency} & \multicolumn{1}{c|}{No}  & Yes \\ \hline
\multicolumn{1}{|l|}{Incident severity}        & \multicolumn{1}{c|}{No}  & No  \\ \hline
\multicolumn{1}{|l|}{Incident cause(s)}        & \multicolumn{1}{c|}{No}  & No  \\ \hline
\multicolumn{1}{|l|}{Physical harm}            & \multicolumn{1}{c|}{No}  & Yes \\ \hline
\multicolumn{1}{|l|}{Environmental harm}       & \multicolumn{1}{c|}{No}  & Yes \\ \hline
\multicolumn{1}{|l|}{Property harm}            & \multicolumn{1}{c|}{No}  & No  \\ \hline
\multicolumn{1}{|l|}{Psychological harm}       & \multicolumn{1}{c|}{No}  & No  \\ \hline
\multicolumn{1}{|l|}{Reputational harm}        & \multicolumn{1}{c|}{No}  & Yes \\ \hline
\multicolumn{1}{|l|}{Economic harm} & \multicolumn{1}{c|}{No}  & Yes \\ \hline
\multicolumn{1}{|l|}{Legal/ regulatory harm}   & \multicolumn{1}{c|}{No}  & Yes \\ \hline
\multicolumn{1}{|l|}{Human rights harm}      & \multicolumn{1}{c|}{No}            & No              \\ \hline
\multicolumn{1}{|l|}{Link to incident description/ news article} & \multicolumn{1}{c|}{No}            & Yes             \\ \hline
\multicolumn{3}{|l|}{\textbf{Redacted fields (submitter details):}}               \\ \hline
\multicolumn{3}{|l|}{Name of submitter}                                         \\ \hline
\multicolumn{3}{|l|}{Email of submitter}                                        \\ \hline
\multicolumn{3}{|l|}{Incident news source(s)}                                   \\ \hline
\multicolumn{3}{|l|}{Extra information shared by the submitter}                 \\ \hline
\end{tabular}
\end{center}
\end{table}
%%%%%%%%%%%

The fields are selected to ensure clarity, consistency, and comprehensiveness, facilitating the accurate collection and analysis of AI incident data across diverse contexts. The description of each field is as follows:
\begin{enumerate}
\item \emph{Incident ID}: A unique identifier assigned to each incident.
\item \emph{Incident Title}: A concise title that encapsulates the incident.
\item \emph{Incident Summary}: A detailed overview of the incident, up to 250 words.
\item \emph{Incident Date}: The exact date (and time, if applicable) when the incident occurred.
\item \emph{Incident Location(s)}: The geographical area(s) where the incident occurred.
\item \emph{Affected Party(ies)}: The individuals, organizations, or entities impacted by the incident.
\item \emph{Sector(s) Impacted}: The industry or sector affected by the incident.
\item \emph{Incident Issue(s)}: The specific concerns related to the system, governance, technology, or third-party actions.
\item \emph{AI Application Name(s)}: The name of the AI system or application involved in the incident.
\item \emph{Application Version}: The specific version of the AI application in use.
\item \emph{Application Technology(ies)}: The technologies employed within the AI application/system.
\item \emph{Application Purpose(s)}: The intended function or goal of the AI application.
\item \emph{Application Deployer}: The organization or entity responsible for deploying the AI system.
\item \emph{Application Developer}: The organization or entity that created the AI system.
\item \emph{Application Transparency}: The clarity, accessibility, and accountability of the AI system to users and stakeholders, including the ability to challenge it.
\item \emph{Incident severity}: The level of impact or seriousness of the incident.
\item \emph{Incident Cause(s)}: The root causes or contributing factors leading to the AI incident.
\item \emph{Physical Harm}: Any form of injury, damage, or adverse impact on the physical well-being of an individual or a group.
\item \emph{Environmental Harm}: Any adverse impact or damage on the environment affecting ecosystems, wildlife, air, water, or soil. 
\item \emph{Property Harm}: Damaging or destroying property of an individual, group, or organization.
\item \emph{Psychological Harm}: Damage to mental health and well-being of an individual or a group.
\item \emph{Reputational Harm}: Damage to the reputation of an individual, group, or organization.
\item \emph{Economic Harm}: Impairment of financial assets of an individual, group, or organization.
\item \emph{Legal/Regulatory Harm}: Any legal or regulatory consequences arising from the incident.
\item \emph{Human Rights Harm}: Damage to fundamental rights or human rights to an individual or a group.
\item \emph{Link to the incident description/ news article}:  A URL directing to external sources for detailed information or news coverage of the incident.
\item \emph{Name of submitter}: The full name of the individual or organization submitting the incident report.
\item \emph{Email of submitter}: The contact email address of the submitter for follow-up and verification purposes.
\item \emph{Incident news source(s)}: The sources, such as news articles or reports, from which information about the incident was obtained.
\item \emph{Extra information shared by the submitter}: Additional details or context provided by the submitter that may enhance the understanding of the incident.
\end{enumerate}
\emph{Note}: Serial numbers 27 to 30 are redacted fields as they pertain to details of the submitter.

\subsection{Proposed taxonomy}
Table \ref{table3} presents the proposed taxonomy for AI incidents in critical digital infrastructure. It is designed to address the unique challenges and impacts of AI failures in areas like telecommunications and energy. A general-purpose taxonomy may not be able to capture the specific nuances and critical aspects of AI incidents in essential sectors. It categorizes incidents by type, affected systems, severity, cause of failure, and harm, providing subcategories and examples such as network disruptions and IoT component failures. This targeted approach ensures that the taxonomy reflects the distinct complexities of critical digital infrastructure, offering detailed insights and actionable data that broader frameworks might overlook. The structured and specialized framework proposed in this work enhances incident analysis and management, contributing to improved resilience and compliance in these crucial sectors.

%\begin{table}[htbp]
\begin{table}[h]
\caption{Taxonomy for critical digital infrastructure database}
\label{table3}
\setlength{\tabcolsep}{1pt}  % reduces spaces in cell padding
\begin{center}
\begin{tabularx}{\columnwidth}
{|
>{\raggedright\arraybackslash}>{\hsize=0.4\hsize}X |
>{\raggedright\arraybackslash}>{\hsize=0.9\hsize}X |
>{\raggedright\arraybackslash}>{\hsize=1.7\hsize}X |
}
\hline
\textbf{Category} &
  \textbf{Subcategory} &
  \textbf{Examples} \\ \hline
\multirow{6}{=}{Incident type}     & Network Disruption               & Telecom network outages, power grid failures.                      \\ \cline{2-3} 
                                   & Service Quality Degradation      & Slower internet speeds, voltage fluctuations.                      \\ \cline{2-3} 
                                   & Security Breach                  & Data breaches, unauthorized access.                                \\ \cline{2-3} 
                                   & AI Mismanagement                 & Incorrect resource allocation, faulty AI decisions.                \\ \cline{2-3} 
                                   & Operational Failure              & Trading system errors, logistics failures.                         \\ \cline{2-3} 
                                   & Predictive Maintenance Failure   & Unpredicted power outages, hardware failures.                      \\ \hline
\multirow{6}{=}{Affected system}  & Core Network & Failure in central telecom switches, energy grid control centers. \\ \cline{2-3} 
                                   & Edge/Access Networks             & Base station disruptions, edge server issues.                      \\ \cline{2-3} 
                                   & Data Transmission Systems        & Data link failures, fiber optic congestion.                        \\ \cline{2-3} 
                                   & Virtualized/Cloud Infrastructure & Cloud service outages, virtual network issues.                     \\ \cline{2-3} 
                                   & IoT Components                   & Faulty smart meters, IoT sensor failures.                          \\ \cline{2-3} 
                                   & Physical Infrastructure          & Security system malfunctions, HVAC failures.                       \\ \hline
\multirow{4}{=}{Incident severity} & Critical                         & Major nationwide outages, complete system failures.                \\ \cline{2-3} 
                                   & High                             & Significant disruptions, major service degradation.                \\ \cline{2-3} 
                                   & Moderate                         & Regional outages, partial service degradation.                     \\ \cline{2-3} 
                                   & Low                              & Minor interruptions, brief service slowdowns.                      \\ \hline
\multirow{4}{=}{Cause of failure} & AI Misconfiguration          & Misconfigured resource settings, faulty automation.               \\ \cline{2-3} 
                                   & Predictive Maintenance Error     & Missed maintenance alerts, undetected failures.                    \\ \cline{2-3} 
                                   & Security Vulnerability           & Exploited AI weaknesses, data breach vulnerabilities.              \\ \cline{2-3} 
                                   & Human-Related AI Errors          & Design flaws, oversight errors.                                    \\ \hline
\multirow{8}{=}{Type of harm}     & Physical Harm                & Injuries from machinery failures, infrastructure damage.          \\ \cline{2-3} 
                                   & Environmental Harm               & Increased emissions, environmental damage.                         \\ \cline{2-3} 
                                   & Property Harm                    & Damage to telecom towers, power substations.                       \\ \cline{2-3} 
                                   & Psychological Harm               & Public anxiety from outages, distress from service disruptions.    \\ \cline{2-3} 
                                   & Reputational Harm                & Loss of trust in service providers, damaged corporate credibility. \\ \cline{2-3} 
                                   & Economic Harm                    & Revenue loss from outages, penalties for non-compliance.           \\ \cline{2-3} 
                                   & Legal/Regulatory Harm            & Fines for GDPR breaches, regulatory sanctions.                     \\ \cline{2-3} 
                                   & Human Rights Harm                & Privacy violations, restricted freedoms from surveillance.         \\ \hline
\end{tabularx}
\end{center}
\end{table}

\section{Discussion}
\label{discussion}
This study aims to address significant gaps in AI incident reporting by introducing a unified schema and taxonomy customized for critical digital infrastructure. Existing databases, such as AIID and AIAAIC, are beset by inconsistent data formats and insufficient granularity, hindering effective incident analysis and integration. Further, they, being general-purpose, lack the specific details required to analyze incidents in specialized sectors, such as critical digital infrastructure.

Our standardized schema directly tackles these issues by incorporating specific fields like incident causes, incident severity, and types of harm. For instance, a general system failure previously reported in AIID is now captured with precise details, such as sectors affected, incident locations, incident issues, etc., allowing for a clearer assessment of impact and contributing factors. This level of detail enhances the ability to analyze and understand the specific nature of each incident, leading to more effective responses and mitigation strategies.

Similarly, AIAAIC's previous reporting often lacked sufficient context, making it difficult to gauge the causes and implications of incidents. Our schema addresses this by including detailed fields. For example, to understand the cause of the incident, the proposed schema includes fields that are not available in AIAAIC, such as incident summary, incident causes, and application version. Similarly, to gauge the impact of an incident, fields missing in AIAAIC have been included to document comprehensive information, such as incident severity, property harm, psychological harm, human rights harm, etc., enabling more targeted and actionable insights.

The proposed taxonomy categorizes incidents by type (e.g., network disruption, security breach), affected systems (e.g., core network, IoT components), severity (e.g., critical, high), causes of failure (e.g., AI misconfiguration, predictive maintenance error), and types of harms (physical, environmental, property, psychological, reputational, economic, legal/regulatory, and human rights harms). This structure facilitates more effective pattern recognition and predictive analysis. For example, incidents involving IoT components are now specifically identified, allowing for the detection of trends and vulnerabilities that were previously hidden under broad classifications.

By standardizing the reporting schema and taxonomy, this work consolidates data from disparate sources into a coherent format, enhancing data clarity and supporting thorough analysis. The improved granularity and detailed categorization not only facilitate better-informed decision-making and policy development but also strengthen resilience and compliance in critical digital infrastructure sectors.

Overall, this unified schema and taxonomy represents a significant advancement in addressing the deficiencies of existing reporting systems, providing a robust framework for systematic documentation, analysis, and management of AI incidents.

\section{Conclusion and future work}
\label{conclusion}
The increasing deployment of AI systems in critical digital infrastructure presents significant potential risks that could lead to system failures or unpredictable operations. The lack of a standardized framework for reporting AI incidents has hindered efforts to collect incident data, systematically analyze them, and develop mitigation strategies. This study addresses this issue by proposing a standardized schema and taxonomy for AI incident databases, enabling more consistent and comprehensive documentation of incidents across various sectors.

One of the primary contributions of this work is the development of a unified schema that enables detailed and structured collection of AI incident data, overcoming previous challenges of integrating disparate databases with incompatible structures. This schema addresses the shortcomings of existing repositories that lacked data fields with the required granularity for a meaningful analysis. Further, incorporating new fields, such as incident severity, causes, and types of harm, facilitates a comprehensive analysis of root causes and diverse impacts. Additionally, the proposed taxonomy enables the systematic classification of AI incidents for further research and policymaking.

This work is crucial as it establishes a foundation for a more unified global response to AI incidents and enables the effective application of lessons learned from one sector or region to others. This work provides significant value across various sectors. For academia and researchers, the standardized schema and taxonomy lay a strong foundation for systematic studies and future research on AI incidents by identifying patterns, predicting incidents, and refining mitigation strategies. Industry can leverage these tools to enhance the safety and reliability of AI systems. For policymakers, the standardized framework serves as a basis for regulatory efforts to ensure the safe and ethical deployment of AI in critical digital infrastructure. By categorizing harms and incident severity, the taxonomy enables regulators to prioritize areas of concern and develop targeted interventions. The detailed, standardized data captured using the proposed schema supports evidence-based policymaking, addressing gaps in existing regulations and frameworks. Lastly, the public benefits from increased transparency and accountability promoted by this standardized approach.

As AI continues to evolve, the standardized schema and taxonomy proposed here will be instrumental in ensuring that AI incidents are systematically documented, analyzed, and addressed in a manner that promotes trust, safety, and accountability in AI systems.

\emph{Future work}: Future efforts should focus on adopting this standardized schema across different sectors and geographies. Additionally, ongoing refinement of the schema, informed by new incidents and emerging technologies, will be crucial. Expanding AI incident reporting and integrating it with automated reporting tools are also key areas for further research.

\bibliographystyle{unsrt}
%\bibliography{references}

\end{document}